\documentclass[12pt]{article}

\usepackage{newtxtext,newtxmath}
\usepackage{graphicx}

\usepackage[letterpaper,margin=1in]{geometry}

\renewenvironment{abstract}
	{\quotation}
	{\endquotation}

\date{}

\makeatletter
\renewcommand{\fnum@figure}{\textbf{Figure \thefigure}}
\renewcommand{\fnum@table}{\textbf{Table \thetable}}
\makeatother

\usepackage{scicite}

\usepackage{url}

\usepackage{comment}
\newcommand\G{\Gamma}
\newcommand\veps{\varepsilon}
\newcommand{\tc}[1]{\textcolor{black}{#1}}

\usepackage{xcolor}
\usepackage{chemformula} % Formula subscripts using \ch{}
\usepackage[T1]{fontenc} % Use modern font encodings
\usepackage{braket}
\usepackage{caption}
\usepackage{subcaption}

\def\scititle{Graphene double quantum dot as a solid-state two-level system in the THz range}
\title{\bfseries \boldmath \scititle}

\author{
Alexis~Wietzke$^{1\dagger}$, Solen~Coeymans$^{1\dagger}$, Benjamin~Hue$^{1}$,\and Simon~Messelot$^{2}$, Aurélien~Schmitt$^{1}$, Michael~Rosticher$^{1}$,\and
Kenji~Watanabe$^{3}$, Takashi~Taniguchi$^{4}$, Jerome~Tignon$^{1}$,\\ Sukhdeep~Dhillon$^{1}$, Rebeca~Ribeiro$^{5}$, Audrey~Cottet$^{1,6}$,\and
Matthieu~Delbecq$^{1}$, Takis~Kontos$^{1,6}$, Juliette~Mangeney$^{1\ast}$\\  % petit espace avant affiliations
\small
$^{1}$Laboratoire de Physique de l’Ecole Normale Supérieure, ENS, Université PSL, CNRS, \\[-4mm] \small Sorbonne Université, Université de Paris, Paris, France\\[-2mm]
\small$^{2}$Quobly, Grenoble, France\\[-2mm]
\small$^{3}$Research Center for Electronic and Optical Materials, National Institute for Materials Science, \\[-4mm] \small 1-1 Namiki, Tsukuba, Japan\\[-2mm]
\small$^{4}$Research Center for Materials Nanoarchitectonics, National Institute for Materials Science, \\[-4mm] \small 1-1 Namiki, Tsukuba, Japan\\[-2mm]
\small$^{5}$Université Paris-Saclay, CNRS, Centre de Nanosciences et Nanotechnologies (C2N), 91120 Palaiseau, France\\[-2mm]
\small$^{6}$Laboratoire de Physique et d’Étude des Matériaux, ESPCI Paris, Université PSL, CNRS, \\[-4mm] \small  Sorbonne Université, Paris, France\\[-2mm]
\small $^\ast$Corresponding author. Email: juliette.mangeney@phys.ens.fr
}

\begin{document} 
% Insert the title and author list
\maketitle

\subsection*{Abstract}
\begin{abstract} \bfseries \boldmath
Matter-based two-level quantum systems are fundamental building blocks of quantum technologies, underpinning applications in quantum computing, simulation, communication, and sensing. Today, most of these two-level quantum systems, including Rydberg atoms, trapped ions, superconducting circuits, molecules, and quantum dots, involve transition frequencies in microwave or optical domains. However, extending their transition frequency to the THz spectral range could provide numerous benefits, leveraging the unique properties of THz radiation. Here, we show that an hBN-encapsulated graphene double quantum dot realizes a two-level system with a transition frequency in the THz spectral range. Using low-temperature transport spectroscopy \tc{in the dark} combined with modeling based on nonequilibrium Green's function formalism, we demonstrate the formation of an artificial molecule, featuring bonding and antibonding states separated by an energy of up to 0.6 meV, corresponding to a resonance frequency of $\sim$0.145 THz. This approach provides direct access to important parameters of \tc{open} double quantum dot systems, including tunneling rates, offering a practical alternative to conventional techniques such as photon-assisted tunneling, which are challenging to implement at THz frequencies. The realized 2D-material-based THz two-level system, compatible with direct on-chip integration, could serve as a key building block for future THz quantum technologies. Furthermore, by involving a single-electron transition, coupling these devices to a THz resonator could enable the exploration of ultra-strong light-matter interactions at the fundamental limit of a single transition coupled to a single cavity, beyond the reach of existing THz quantum systems based on collective transitions.
\end{abstract}

% The first paragraph of any Science paper does NOT have a heading
% Nor is it indented
\noindent

\subsection*{Introduction}

Quantum technologies have advanced rapidly at microwave and optical frequencies. By contrast, their extension into the terahertz (THz) range remains in its infancy, as referred to as the THz quantum gap, despite its significant technological promise \cite{Todorov2024, Vitiello2022, Pontula2025, Zhang2025}. For example, THz radiation is highly attractive for secure wireless communications due to its low sensitivity to atmospheric disturbances \cite{Liu2025}, for quantum computing with its potential to increase the operating temperature of solid-state qubits, which would address scalability issues, and for quantum sensing owing to its ability to penetrate otherwise opaque materials \cite{Mirco2025}. A key milestone in bridging the THz quantum gap is to realize two-level systems with transitions at THz frequencies, as these systems can serve as qubits, and when coupled to resonators, enable the generation and detection of non-classical light states, such as single photons, squeezed light, Fock states, and entangled photons \cite{Walls1983, Andersen_2016, Vahlbruch2016,Castellanos2008}. Vibrational or rotational states in several molecules\cite{Najafian2020, Wang2022, Wang2023} and highly excited Rydberg states in atoms \cite{Wade2017, Nill2024} can be harnessed as two-level systems at THz frequencies, but both approaches face limitations in scalability and operational complexity, primarily due to the requirement for isolated environments or trapping schemes. \tc{Another candidate for a THz two-level system is the 1s–2p+ donor transition in bulk GaAs under magnetic field. However, the 2p+ state lying within the conduction band leads to strong decoherence, limiting coherent control of the transition \cite{cole2001}.} Developing a novel technology to realize a solid-state two-level quantum system at THz frequencies compatible with direct on-chip integration would represent a major breakthrough that could considerably accelerate the progress and deployment of THz quantum technologies.\\

Single quantum dots with characteristic sizes of a few tens to a few hundreds of nanometers are appealing for implementing THz two-level systems as they exhibit multiple dipolar transitions at THz frequencies, as recently observed in various material systems, including carbon nanotubes \cite{Valmorra2021},  InAs/InAs$_{0.3}$P$_{0.7}$ nanowires \cite{Asgari2021}, AlGaAs/GaAs two-dimensional electron gas \cite{Kuroyama2024}, colloidal HgTe nanocrystals \cite{Apretna2021} and graphene \cite{El_Fatimy2016, Riccardi2020}. Among these, graphene quantum dots are particularly attractive, featuring substantial THz electric dipole moments (d $\approx$ 230 nm)\cite{Messelot2022} and straightforward on-chip integration thanks to the graphene two-dimensional structure. However, realizing a two-level system in a single quantum dot, which is widespread in the optical domain \cite{Kasprzak2010}, faces a major challenge at THz frequencies due to the difficulty of isolating the two-level system from other multiple electronic states. Indeed, in this low-energy regime, broadening effects become comparable to transition energies, leading to blurring of the energy differences between multiple transitions. This challenge is effectively addressed in a double quantum dot (DQD), where the transition energy of the two-level system can be isolated from other transition energies via tuning the interdot tunnel coupling \cite{Petta2005}. \tc{Such tunability is also crucial for controlling and manipulating the two-level system, yet it is absent in a single QD, where the electronic level spectrum is determined by the spatial confinement of the electrons and the corresponding transition energies are fixed during device fabrication.} Although DQD systems have been extensively investigated in the microwave regime \cite{Molitor_2010, Liu2010, Banszerus2020, Dani2018, Liu_2014, Lundberg2024, Denisov_2022, vanderWiel_2001, Neukelmance2025, Banszerus_tunable_coupling_BLG_QD}, where two-level systems with resonance frequencies up to 50 GHz have been demonstrated \cite{Zhang_2023}, their realization in the THz spectral range remains \tc{to a great extent unexplored \cite{wang2012}}.\\

In this work, we demonstrate a two-level system in an hBN-encapsulated graphene DQD with a resonance frequency of up to $\sim$0.145 THz. By combining low-temperature DC transport spectroscopy \tc{in the dark} with modeling based on nonequilibrium Green's function formalism using equation-of-motion method, we quantitatively capture the relevant energy scales of the \tc{open} two-level system and its coupling to the reservoirs. \tc{These results highlight the potential of transport spectroscopy in the dark as a tool for probing open two-level systems in the THz regime}, where conventional approaches such as photon-assisted tunneling are challenging to implement. \tc{Here, the term “THz two-level system” refers to a two-level system with a transition energy in the THz range, rather than to a system actively driven by THz radiation.} From a technological perspective, the hBN-encapsulated graphene DQD, fully compatible with on-chip integration, could serve as a promising building block for THz quantum devices, potentially capable of bridging the current THz quantum gap. Moreover, involving a single-electron transition, this system could, when coupled to a THz resonator, provide a platform to explore ultra-strong light–matter interactions at the fundamental limit of a single quantum transition coupled to a single cavity, addressing a long-standing challenge in quantum optics. \\

\subsection*{Results}
\begin{figure}[!ht]
    \makebox[\textwidth][c]{% to center
    \includegraphics[width=18.4cm]{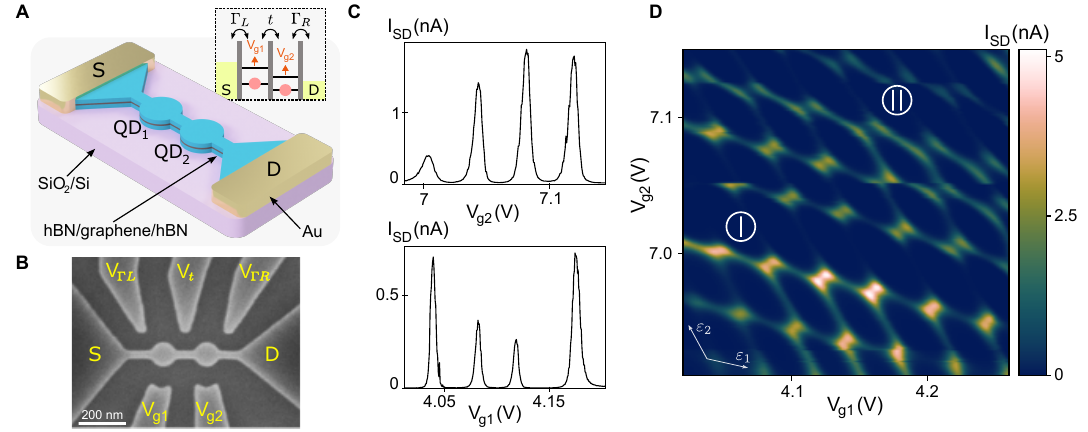}}
    \caption{\textbf{Artificial molecule made of hBN-encapsulated graphene DQD.} (\textbf{A}) Schematic illustration of the device architecture: a DQD based on hBN/graphene/hBN heterostructure in a single electron transistor geometry (bottom). Schematic of the energy levels of the two QDs and the electrochemical potentials of the source and drain reservoirs for positive bias (top). Quantum transport through the device involves tunneling from source to the left QD with a $\Gamma_L$ energy, between the two QDs with a $t$ energy and from the right QD to drain with a $\Gamma_R$ energy. (\textbf{B}) A false-color scanning electron microscopy image of the hBN-encapsulated graphene section of the device. The 130 nm-diameter QDs and 40 nm-width constrictions are capacitively coupled to lateral gates. (\textbf{C}) Top (bottom): Current as a function of V$_{g2}$ (V$_{g1}$) at V$_{DS}$=0.465 mV, exhibiting Coulomb blockade peaks. During the sweep, V$_{g1}$ (V$_{g2}$) is simultaneously varied from 4.01 to 4.20 V (from 6.99 V to 7.15 V) to follow the direction $\veps_1$ ($\veps_2$) indicated in \textbf{D}, along which only the chemical potential of the left (right) QD varies. (\textbf{D}) Charge stability diagram of the graphene DQD represented by $I_{SD}$ as a function of V$_{g1}$ and V$_{g2}$,  recorded over a gate-voltage range of approximately $\pm$0.1 V around 4.15 V and 7.05 V, respectively, at V$_{DS}$=0.465 mV and V$_{\text{t}}$= -0.27 V. In this gate-voltage range, different coupling regimes between the two QDs are observed: intermediate coupling (region I) and strong coupling (region II).}
    \label{fig:1}
\end{figure}

The studied devices consist of an hBN-encapsulated graphene DQD in a single-electron transistor geometry (Fig.~\ref{fig:1}A). The fabrication process involves the sequential stacking of mechanically exfoliated hBN and graphene flakes to form an hBN/graphene/hBN heterostructure, assembled using a standard dry van der Waals transfer technique \cite{Wang2013}. The heterostructure is transferred onto a high-resistivity silicon substrate ($>8000$ ohms.cm) covered with a 285 nm thick thermal oxide (SiO$_{2}$). Encapsulating graphene in hBN significantly reduces bulk disorder and preserves graphene from chemical contamination \cite{Novoselov2016,Banszerus2015,Engels2013}. Using electron-beam lithography followed by CHF$_{3}$ and O$_{2}$-based reactive ion etching on the hBN/graphene/hBN heterostructure, we define two 130 nm-diameter graphene QDs connected to each other and to source and drain electrodes by narrow constrictions (width $\sim$40 nm) that act as tunnel barriers (Fig.~\ref{fig:1}B). The QDs and constrictions are capacitively coupled to lateral gates, also made of hBN-encapsulated graphene.  Using a second electron-beam lithography and metal evaporation followed by lift-off, we then connect the source, drain  electrodes of the QDs and lateral gates to Cr/Au electrodes using 1D edge-contact techniques \cite{Wang2013} (see Materials and methods for further details). In our device architecture, the electrochemical potentials of the left and right QD are mainly adjusted by the plunger gate voltages V$_{g1}$ and V$_{g2}$, respectively. The tunneling energies from source to the left QD ($\G_L$), between the two QDs ($t$) and from the right QD to drain ($\G_R$) are tuned using gate voltages V$_{\Gamma \text{L}}$, V$_{\text{t}}$ and V$_{\Gamma \text{R}}$, respectively \cite{Guttinger2012}. Applying a bias voltage V$_{DS}$ shifts the source electrochemical potential relative to the Fermi energy, opening a bias window through which current flows when the QD electrochemical potentials align to form a downward pathway (Fig.~\ref{fig:1}A). We probe the quantum transport in the hBN-encapsulated graphene DQD devices using low-temperature transport spectroscopy, mounting the devices in a $^{3}$He cryostat with a base temperature of 280 mK and using DC measurement techniques (see Methods for the measurement details). \\

We initially set transport through the graphene DQD device in the Coulomb blockade regime by applying V$_{\G \text{L}}$=1.88 V and V$_{\G \text{R}}$=1.15 V, ensuring that the constrictions between the QDs and the leads act as tunneling barriers for electrons. This regime is evidenced by the pronounced Coulomb blockade peaks observed in the tunnel current, $I_{SD}$, as a function of the gate voltages V$_{g1}$ and V$_{g2}$ (Fig.~\ref{fig:1}C). We then study the different transport regimes by recording charge-stability diagrams. Figure \ref{fig:1}D shows the tunneling current $I_{SD}$ as a function of V$_{g1}$ and V$_{g2}$, measured over a gate-voltage range of about $\pm$0.1 V centered around 4.15 V and 7.05 V, respectively, at V$_{DS}$=0.465 mV and V$_{\text{t}}$= -0.27 V. Although the gate voltages V$_{g1}$ and V$_{g2}$ primarily tune the chemical potentials of the QDs, they also noticeably modulate their mutual coupling via cross-capacitance effects \cite{Liu2010}, thereby modifying the transport regime. In region I, the charge stability diagram exhibits the honeycomb pattern characteristic of a DQD in the intermediate coupling regime with the cell pattern repeated several times \cite{Banszerus2020}. Elastic charge transport through the graphene DQD occurs only at the vertices of the honeycomb pattern, where the charge states of both QDs align within the source-drain bias window. At these vertices, the tunneling current forms pairs of triangles, the finite-bias counterparts of the zero-bias triple points, where three regions of stable charge occupancy intersect \cite{vanderWiel2002}. The edges of the honeycomb cells are visible due to off-resonant current, which originates from co-tunneling processes \cite{Gustavsson2008}. These occur when the electrochemical potential of one QD aligns with that of its neighboring lead, allowing charge transport via an intermediate virtual state \cite{Gustavsson2008}. In region II, the increased separation between triangles indicates stronger interdot coupling, reaching the strong coupling regime and leading to a transition of the device towards a single large QD \cite{Livermore1996}. \\

\begin{figure}[hht!]
    \centering
    \makebox[\textwidth][c]{% to center
    \includegraphics[width=18.4cm]{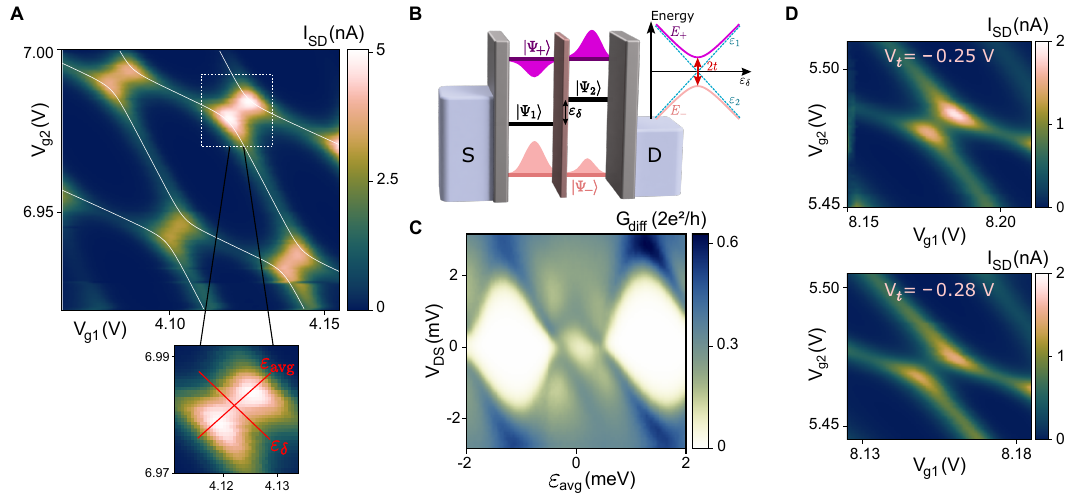}}
    \caption{\textbf{Probing the DQD device in the intermediate coupling regime.} (\textbf{A}) Charge stability diagram in the intermediate interdot coupling regime. White lines represent tunneling current lines predicted by modeling the DQD system with a Hamiltonian that includes both the purely electrostatic coupling between the QDs and a finite interdot tunnel coupling $t$ (see Supplementary for details). (\textbf{B}) Left: schematic diagram of a two-level system with unperturbed eigenstates $\ket{\Psi_1}$ and $\ket{\Psi_2}$ and eigenstates belonging to the symmetric (bonding) state, $\ket{\Psi_-}$, and the anti-symmetric (antibonding) state, $\ket{\Psi_+}$. Right: Energies $E_-$ and $E_+$ versus the energy difference $\varepsilon_\delta$. (\textbf{C}) Differential conductance $G_{\text{diff}}$ as a function of the source-drain voltage $V_{DS}$ and of the mean energy $\veps_{\text{avg}}$ at $\varepsilon_\delta$=0 and V$_\text{t}$=-0.246 V, revealing a small secondary Coulomb blockade diamond in between two large Coulomb diamonds. (\textbf{D}) Close view of the interdot coupling represented by the triangles splitting with tuning V$_\text{t}$ from -0.25 V to -0.28 V.}
    \label{fig:2}
\end{figure}

We now focus on the intermediate interdot coupling regime and analyze quantitatively the charge stability diagram, shown in Fig.~\ref{fig:2}A. Using an electrostatic model that accounts for cross-capacitances between opposite QDs and plunger gates \cite{vanderWiel2002}, we determine the electrostatic energies of the graphene DQD device from the dimensions of the triangles at the vertices of the honeycomb cell and the separation between their centers. The model yields the lever arms, the charging energies of left and right QDs,  $E_{CL}=1.83$ meV and $E_{CR}=2.54$ meV respectively, as well as their mutual charging energy, $E_m=0.41$ meV (see Supplementary Information for details). The charging energies and lever arms of each QD are relatively close, indicating a high degree of uniformity in the QD sizes and their distances from the plunger gates. No clear resonant line is observed at the base of the triangles, consistent with the large single-particle level spacing ($\sim 1.7$ meV) in graphene QDs, \tc{which is larger than $V_{DS}$} \cite{Riccardi2020}. \\

Notably, the co-tunneling lines in the charge stability diagram do not intersect at the triangle centers (local maxima) and exhibit a parabolic curvature near the triangle edges, a signature of energy-level hybridization \cite{vanderWiel2002}. This hybridization arises from coherent electron tunneling between the coupled QDs, resulting in delocalization of the electron wavefunctions across the DQD. These delocalized states allow the flow of current through co-tunneling processes, giving rise to current bands that supplement the triangular current region associated with the sequential regime. At low energy detuning between the QD levels, the strong hybridization produces bonding and antibonding molecular-like states, which together form a two-level system, as illustrated in Fig.~\ref{fig:2}B. At zero energy detuning, the resonance frequency of the two-level system is determined by the interdot coupling energy $t$, following $f_{\text{res}}=2t/h$. By modeling the DQD system with a Hamiltonian that includes both the purely electrostatic coupling between the QDs and a finite interdot tunnel rate $t$ (see Supplementary for details), we accurately reproduce the co-tunneling current lines, which follow the eigenenergies of the delocalized electronic states for different charge configurations (white lines in Fig.~\ref{fig:2}A). The model captures the observed anti-crossing features and yields a rough estimate of the tunnel coupling energy, $t \sim 0.15 \pm 0.05$ meV. The good agreement between experiment and theory across multiple charge states validates our methodology for extracting the lever arms, i.e., the conversion between gate voltage and energy. We then define two perpendicular axes in the charge stability diagram, highlighted by red lines in the inset of Fig.~\ref{fig:2}A: one along the energy detuning between the states of the two QDs, $\varepsilon_\delta = \mu_{1} - \mu_{2}$, and the other along the average energy $\veps_{\text{avg}} = (\mu_{1} + \mu_{2})/2$, oriented parallel to the baseline of the triangles. \\

Coupling between the QDs is further revealed by measurements of the differential conductance as a function of $V_{DS}$ and the average energy $\veps_{\text{avg}}$, which controls the electrochemical potential of the artificial molecule as a whole. The conductance colormap in Fig.~\ref{fig:2}C displays two large primary Coulomb blockade diamonds,  with half-heights of $\sim$2.2 meV, separated by a smaller secondary diamond with a half-height of 1 meV. While the large diamonds arise from charge blockade in the individual QDs due to their charging energies \cite{Riccardi2020}, the smaller diamond reflects interdot coupling \cite{Crouch1997}. This coupling between the two QDs can be further finely controlled by means of V$_{\text{t}}$ tuning in our device architecture. As shown in Fig.~\ref{fig:2}D, decreasing V$_{\text{t}}$ by 30 mV leads to a slight increase in the separation between the triangles, reflecting an enhanced charging energy and interdot tunnel coupling.\\

\begin{figure}[!ht]
    \centering
    \makebox[\textwidth][c]{% to center
    \includegraphics[width=18.4cm]{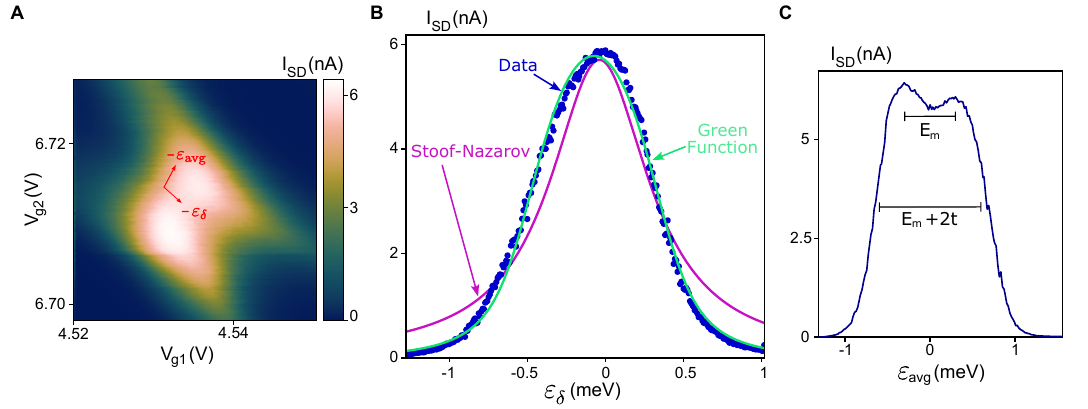}}
    \caption{\textbf{Determining the interdot tunnel rate of the two-level system.} (\textbf{A}) High-resolution charge stability diagram of a single pair of triangles measured over a gate-voltage region centered around V$_{g1}$=4.536 V and $V_{g2}$=6.713 V, at $V_{\text{t}}$=-0.27 V and $V_{DS}$=0.66 mV. (\textbf{B}) Line cut of the tunneling current $I_{SD}$ along the detuning energy axis $\varepsilon_\delta$ at small energy detuning. Blue dots represent experimental data, the purple line is a fit using the Stoof and Nazarov (Eq.\ref{stoof_naravo_equation}), and the green line corresponds to a fit using nonequilibrium Green's function formalism. (\textbf{C}) Line cut of the tunneling current $I_{SD}$ along the mean energy axis $\veps_{\text{avg}}$. We set $\veps_{\text{avg}}$=0 at the center of the triangle separation (i.e. current-band separation). In this coupling regime, $E_m$=0.60 meV and $t$=0.30 meV.}
    \label{fig:3}
\end{figure}

The interdot coupling energy $t$ sets the order of magnitude of the transition frequency of the two-level system. To quantify it, we analyze the tunneling current along the energy detuning axis, where the  average total charge of the DQD is constant. To this end, we record a high-resolution charge stability diagram of a single pair of triangles, shown in Fig.~\ref{fig:3}A, measured over a different gate-voltage region centered around V$_{g1}$=4.536 V and $V_{g2}$=6.713 V, where the coupling regime results in an enhanced state hybridization. The mutual charging energy in this regime is $E_m$=0.60 meV (see Supplementary). Figure \ref{fig:3}B displays a cut of the tunneling current $I_{SD}$ along the detuning energy $\varepsilon_\delta$ axis (blue dots) taken at $\veps_{\text{avg}}$ corresponding to the center of the triangle separation (i.e. at the center of the current-band separation), revealing a broad resonance peak as the energy states of the two QDs align. The current profile is noticeably asymmetric, with a sharper front on the positive detuning side (right side of the peak) than on the negative detuning side (left side of the peak). Transport through the ground states of two coupled QDs is routinely analyzed quantitatively using the Stoof and Nazarov equation, derived in the sequential tunneling limit\cite{Stoof_Nazarov_1996, Nazarov_1993}. This model describes resonant tunneling of Coulomb-interactive electrons when only elastic tunneling is important and assuming that $eV_{DS}>k_{B}T,\varepsilon_\delta$ and $eV_{DS}<E_{C1,2}$. In this limit, the source-drain current $I_{SD}$ as a function of the detuning energy $\varepsilon_\delta $ was obtained exactly by Stoof and Nazarov. It exhibits a Lorentzian line shape, expressed as:

\begin{equation} \label{stoof_naravo_equation}
    I_{SD}(\varepsilon_\delta) = \frac eh \frac{\Gamma_R  t^2}{ \varepsilon_\delta^2 + \frac{\Gamma_R^2}{4} + t^2(2 + \frac{\Gamma_R}{\Gamma_L})}
\end{equation}

where $h$ is the Planck constant, $\Gamma_L$ ($\Gamma_R$) the tunneling energies from source (drain) to QD$_1$(QD$_2$). The resonance width is only determined by the lifetime of the QD energy states and is independent of the electron temperature. Since this model neglects contributions from excited states that can occur at negative detuning and disregards inelastic tunneling processes, which become important at negative detuning \cite{Liu2010, vanderWiel2002}, we restrict the fit of Eq.\ref{stoof_naravo_equation} essentially to the right side of the current peak (purple line on Fig.~\ref{fig:3} B). We observe that the model underlying Eq.\ref{stoof_naravo_equation} is not able to account for the experimental data, as the measured $I_{SD}$ curve exhibits a pronounced deviation from a Lorentzian line shape. One interesting possibility is that the sequential tunneling limit, which requires $\Gamma_L,\Gamma_R<k_{B}T$, is in practice very difficult to meet in experiments. In our case, as in many studies on DQD in the GHz range \cite{Banszerus_tunable_coupling_BLG_QD, Fujisawa1998,Banszerus2020,vanderVaart1995}, the thermal energy $k_{B}T$ ($\sim$25 $\mu$eV) is much smaller than the width of the tunneling current resonance, calling for a more elaborate theoretical approach to analyze our data. Nevertheless, using the simplified theoretical approach described by Eq.\ref{stoof_naravo_equation}, we can estimate an interdot coupling energy of $t$=0.22 meV, as well as dot-lead tunnel coupling energies of $\Gamma_R$=0.44 meV and $\Gamma_L\gg \Gamma_R$. This yields a resonance frequency of $f_{\text{res}}=2t/h=$0.107 THz for the two-level system formed by the graphene DQD. \\

For a more accurate description of our experimental data, we model the transport through the DQD using nonequilibrium Green's function based on the equation-of-motion method and the Meir–Wingreen current formula \cite{Meir1992}.  This approach, widely used to investigate the spectral properties of electrons in DQD systems \cite{Sztenkiel_2007, Bulka2004, Lamba2000, You1999}, has been only rarely applied to transport measurements in the GHz range and remains largely unexplored in the THz range. Importantly, this approach remains valid for $\Gamma_L, \Gamma_R > k_B T$, thereby extending beyond the sequential tunneling limit \tc{and enabling physical insights into open DQD systems through a quantitative determination of the interdot coupling energy}. We restrict our analysis to a single spinless energy level per QD, neglect electronic correlations and inter- and intra-dot Coulomb repulsion energies to construct the retarded Green's function $\mathbf{G}^r(\omega)$ between the two QDs as: 
\begin{equation}
G^r_{12}(\omega) =\frac{t}{(\omega-\veps_1+i\G_L/2)(\omega-\veps_2+i\G_R/2)-t^2}
\label{Gr}
\end{equation}
where $\veps_i$ is the single-particle on-site energy  on QD $i$ and $\Gamma_{i}$ is the tunneling energies from the QD to its corresponding leads, in the usual wide-band approximation for the leads spectral density. At finite temperature, the linear current is given by the expression:

\begin{equation}
    I_{SD}=\frac{2e}{\hbar}\int\frac{d\omega}{2\pi}(f_L(\omega)-f_R(\omega))\frac{\G_L\G_Rt^2}{\vert(\omega-\veps_1+i\G_L/2)(\omega-\veps_2+i\G_R/2)-t^2\vert^2}
\end{equation}

where $f_l(\omega)$ denotes the Fermi-Dirac distribution function for the $l$ electrode. This integral can be conveniently re-expressed into a compact expression (see Supplementary for calculation details):

\begin{equation}
    I_{SD}=\frac{2e}{h}\G_L\G_R t^2\sum_{\eta=\pm 1}\mathfrak{Re}\left[Z^{LR}_\eta/ K_\eta\right]
    \label{ISD_fin}
\end{equation}

with $Z^\mu_\eta=\Psi\left(\frac{1}2+i\frac{\mu-(\veps_{\text{avg}}+i\G_{\text{avg}}/2+\eta.r)}{2\pi k_BT}\right)$, $ Z^{LR}_\eta=Z^{\mu_L}_\eta-Z^{\mu_R}_\eta$, $\G_{\text{avg}}=\frac{\G_L+\G_R}2$, $ \veps_{\text{avg}}=\frac{\veps_1+\veps_2}2$,\newline $ K_\eta=\eta. r^*.[(\eta .r^*+i\G_{\text{avg}})^2-r^2]$ and $ r=\frac12\sqrt{\left((\veps_1-\veps_2)-\frac i2(\G_L-\G_R)\right)^2+4t^2}$. The experimental tunneling current over the full range of detuning is very well reproduced by the model, green line in Fig.~\ref{fig:3}B. In particular, the model captures the sharp suppression of the current at large negative and positive detuning, which deviates from a simple Lorentzian lineshape. The best agreement with theory is obtained for an interdot coupling energy of $t$=0.30 $\pm 0.025$ meV, corresponding to a resonance frequency of the graphene DQD two-level system $f_{\text{res}}=2t/h=$0.145 $\pm 0.012$ THz, thereby operating in the THz spectral range. This resonance frequency is approximately three times higher than the values reported to date for two-level systems based on condensed-matter platforms \cite{Molitor_2010, Liu2010, Banszerus2020, Dani2018, Liu_2014, Lundberg2024, Denisov_2022, vanderWiel_2001, Neukelmance2025}. \\

We also extract from the fit the dot-lead tunnel coupling energies of $\Gamma_L$=0.04 meV and $\Gamma_R$=0.28 meV, which confirms that the thermal energy is smaller than the width of the energy levels of the isolated QDs, and thus not in the limit for applying Eq.\ref{stoof_naravo_equation}. The ability of our model, which accounts only for elastic tunneling processes, to quantitatively reproduce the experimental data at large negative detuning suggests that the current is not limited by inelastic interdot tunneling. This could indicate that $\Gamma_L$ and $\Gamma_R$ exceed the inelastic tunneling rate associated with the emission of energy quanta, such as phonons or photons \cite{Tarucha1999}. Note also that, in order to be consistent with the existing literature \cite{Stoof_Nazarov_1996, Nazarov_1993}, we have used a spinless theory. In our case, this leads to a factor of 4 change in the current expression, which can be absorbed in the $\Gamma$'s. Our approach provides a powerful method for determining the tunneling rates in such quantum systems, overcoming the technical limitations of photon-assisted tunneling spectroscopy, a technique widely used to probe two-level systems at microwave frequencies \cite{vanderWiel_2001, Hecker2023, Kouwenhoven1999}. In photon-assisted tunneling spectroscopy, electrons are driven across the interdot barrier by microwave excitation; however, extending this approach to the THz range is challenging due to strong transmission-line losses at THz frequencies \cite{Navarro2016}, typically requiring the use of dilution cryostats with optical access. Figure \ref{fig:3}C reports the tunneling current along the average energy axis at weak energy detuning. As the average energy is varied, changes in the charge configuration give rise to two current maxima corresponding to the centers of the triangles, separated by the mutual charging energy $E_m$. In addition, the coherent charge tunneling between the discrete levels of the two QDs, which leads to the formation of bonding and antibonding states, is expected to contribute to the broadening of the tunneling current trace. Consistent with this picture, the measured half-width at half maximum of the current trace closely matches $E_m+2t$=1.2 meV. Further theoretical developments incorporating interdot interactions \cite{Sobrino2024} should enable quantitative interpretation of this experimental trace.  Our model thus provides key insights into transport through the hybridized states of the DQD, including occupation numbers and tunneling currents and yields explicit functional dependencies that are essential for device design and for exploring new physical regimes. \tc{Its applicability to study open DQD systems offers an important extension beyond existing methods, which are largely limited to closed DQD systems that are difficult to implement.}\\

\subsection*{Discussion}
In conclusion, we have realized a two-level system operating in the THz spectral range based on a DQD formed in an hBN/graphene/hBN heterostructure. By combining low-temperature quantum transport spectroscopy \tc{in the dark} with a nonequilibrium Green’s function formalism based on the equation-of-motion method, we have extracted important parameters of the \tc{open} two-level system, namely the interdot coupling energy and the tunneling energies with reservoirs. Our work demonstrates that the 2D material-based two-level system exhibits a resonance frequency as high as 0.145 THz.
\tc{Our approach is applicable to open DQD systems, thereby circumventing the experimentally challenging sequential tunneling regime,} and establishes a practical experimental and theoretical framework for studying  quantum systems in the terahertz range using only transport spectroscopy \tc{in the dark}. From a technological perspective, this on-chip-compatible THz two-level system, featuring a large electric dipole moment ($d\approx 230$ nm) \cite{Messelot2022} and straightforward integration with state-of-the-art THz resonators, has the potential to bridge the current THz quantum gap and could provide a platform for the generation and detection of non-classical THz light. Regarding the fundamental implications of our THz DQD system, which involves a single-electron transition rather than an ensemble, coupling it to a THz resonator with extremely small mode volumes \cite{Kellert2017, Messelot2023} could enable the exploration of ultra-strong (deep) light–matter interactions at the ultimate limit of a single quantum transition coupled to a THz resonator, a central challenge in quantum optics \cite{Baydin2025, Johl2025}. Moreover, extending our theoretical framework beyond the wideband approximation and to higher orders in perturbation opens the possibility of investigating correlations with conductors, inter- and intra-dot interactions, and many-body phenomena such as the Kondo effect in a graphene-based DQD.\\

\subsection*{Materials and methods}
\subsubsection*{Device fabrication}
The hBN/graphene/hBN heterostructure is fabricated starting with the micromechanical exfoliation of monolayer graphene and hBN flakes. Afterwards, we compose the heterostructure by using the hot pick-up technique and transfer it onto an intrinsic silicon (Si) substrate with a 285-nm-thick thermal oxide (SiO$_{2}$) layer on top. The Si substrate has a resistivity $>$ 8000 ohms.cm and the small residual N-type doping is due to phosphorus impurities. Atomic force microscopy measurements indicate a bottom hBN thickness of $\sim$10 nm and a top layer of $\sim$50 nm. Using Hydrogen Silsesquioxane (HSQ) negative resist, electron-beam lithography (EBL) followed by Reactive Ion Etching with CHF$_3$ and O$_2$ as agents, we define the double quantum dot (DQD)-based single electron transistor (see  Fig. \ref{fig:1}B). The device consists of two QDs of diameter $\sim$ 130 nm connected to each other and to the source and drain reservoirs via three constrictions of width $\sim$ 40 nm, and five side gate electrodes. A second EBL step is performed to connect the graphene DQD-based device to gold electrodes via 1D edge contact \cite{Wang2013}, in which conduction occurs along the atomic edges of the graphene. To minimize contact resistance, we design a comb-shaped structure that maximizes the length of the 1D contact.

\subsubsection*{Charge stability diagram analysis}
We convert the plunger-gate voltages into quantum dot energies using a capacitance model of the double quantum dot system. This model explicitly includes both the mutual charging energy between the dots and the cross-capacitances between each dot and the opposite plunger gate. We incorporate the interdot tunnel coupling by constructing the Hamiltonian in the charge-state basis and diagonalizing it within fixed-charge subspaces. This procedure enables us to compute the hybridized eigenenergies and reproduce the experimentally observed anti-crossings in the stability diagram. A detailed description of the method is provided in the Supplementary Materials.

%%%%%%%%%%%%%%%% MAIN TEXT FIGURES %%%%%%%%%%%%%%%

%%%%%%%%%%%%%%%% REFERENCES %%%%%%%%%%%%%%%

\clearpage 
\bibliography{biblio}

@article{Todorov2024,
title = {THz quantum gap: exploring potential approaches for generating and detecting non-classical states of THz light},
author = {Yanko Todorov and Sukhdeep Dhillon and Juliette Mangeney},
pages = {1681--1691},
volume = {13},
number = {10},
journal = {Nanophotonics},
doi = {10.1515/nanoph-2023-0757},
year = {2024},
lastchecked = {2024-11-12}
}

@article{Vitiello2022,
author = {Vitiello, Miriam Serena and De Natale, Paolo},
title = {Terahertz Quantum Cascade Lasers as Enabling Quantum Technology},
journal = {Adv. Quantum Technol.},
volume = {5},
number = {1},
pages = {2100082},
doi = {10.1002/qute.202100082},
year = {2022}
}

@article{Walls1983,
	author = {Walls, D.  F. },
	date = {1983/11/01},
	doi = {10.1038/306141a0},
	id = {Walls1983},
	isbn = {1476-4687},
	journal = {Nature},
	number = {5939},
	pages = {141--146},
	title = {Squeezed states of light},
	volume = {306},
	year = {1983}}

@article{Andersen_2016,
doi = {10.1088/0031-8949/91/5/053001},
year = {2016},
month = {apr},
publisher = {IOP Publishing},
volume = {91},
number = {5},
pages = {053001},
author = {Ulrik L Andersen and Tobias Gehring and Christoph Marquardt and Gerd Leuchs},
title = {30 years of squeezed light generation},
journal = {Phys. Scr.},
}

@article{Vahlbruch2016,
  title = {Detection of 15 dB Squeezed States of Light and their Application for the Absolute Calibration of Photoelectric Quantum Efficiency},
  author = {Vahlbruch, Henning and Mehmet, Moritz and Danzmann, Karsten and Schnabel, Roman},
  journal = {Phys. Rev. Lett.},
  volume = {117},
  issue = {11},
  pages = {110801},
  numpages = {5},
  year = {2016},
  month = {Sep},
  publisher = {American Physical Society},
  doi = {10.1103/PhysRevLett.117.110801}
}

@article{Castellanos2008,
	author = {Castellanos-Beltran, M. A. and Irwin, K. D. and Hilton, G. C. and Vale, L. R. and Lehnert, K. W.},
	date = {2008/12/01},
	doi = {10.1038/nphys1090},
	id = {Castellanos-Beltran2008},
	isbn = {1745-2481},
	journal = {Nat. Phys.},
	number = {12},
	pages = {929--931},
	title = {Amplification and squeezing of quantum noise with a tunable Josephson metamaterial},
	volume = {4},
	year = {2008}
}

@article{Wang2022,
author = {Likun Wang  and Yunpeng Xia  and W. Ho },
title = {Atomic-scale quantum sensing based on the ultrafast coherence of an H$_2$ molecule in an STM cavity},
journal = {Science},
volume = {376},
number = {6591},
pages = {401-405},
year = {2022},
doi = {10.1126/science.abn9220}
}

@article{Wang2023,
  title = {Electrical Manipulation of Quantum Coherence in a Two-Level Molecular System},
  author = {Wang, Likun and Bai, Dan and Xia, Yunpeng and Ho, W.},
  journal = {Phys. Rev. Lett.},
  volume = {130},
  issue = {9},
  pages = {096201},
  numpages = {6},
  year = {2023},
  month = {Feb},
  publisher = {American Physical Society},
  doi = {10.1103/PhysRevLett.130.096201}
}

@Article{Najafian2020,
author = {Najafian, Kaveh and Meir, Ziv and Willitsch, Stefan},
title  = {From megahertz to terahertz qubits encoded in molecular ions: theoretical analysis of dipole-forbidden spectroscopic transitions in N2+},
journal  = {Phys. Chem. Chem. Phys.},
year  = {2020},
volume  = {22},
issue  = {40},
pages  = {23083-23098},
publisher  = {The Royal Society of Chemistry},
doi  = {10.1039/D0CP03906C}
}

@article{Valmorra2021,
	author = {Valmorra, F. and Yoshida, K. and Contamin, L. C. and Messelot, S. and Massabeau, S. and Delbecq, M. R. and Dartiailh, M. C. and Desjardins, M. M. and Cubaynes, T. and Leghtas, Z. and Hirakawa, K. and Tignon, J. and Dhillon, S. and Balibar, S. and Mangeney, J. and Cottet, A. and Kontos, T.},
	date = {2021/09/16},
	doi = {10.1038/s41467-021-25733-x},
	id = {Valmorra2021},
	isbn = {2041-1723},
	journal = {Nat. Commun.},
	number = {1},
	pages = {5490},
	title = {Vacuum-field-induced THz transport gap in a carbon nanotube quantum dot},
	volume = {12},
	year = {2021}}

@article{Asgari2021,
author = {Asgari, Mahdi and Coquillat, Dominique and Menichetti, Guido and Zannier, Valentina and Diakonova, Nina and Knap, Wojciech and Sorba, Lucia and Viti, Leonardo and Vitiello, Miriam Serena},
title = {Quantum-Dot Single-Electron Transistors as Thermoelectric Quantum Detectors at Terahertz Frequencies},
journal = {Nano Lett.},
volume = {21},
number = {20},
pages = {8587-8594},
year = {2021},
doi = {10.1021/acs.nanolett.1c02022}
}

@article{Kuroyama2024,
  title = {Coherent Interaction of a Few-Electron Quantum Dot with a Terahertz Optical Resonator},
  author = {Kuroyama, Kazuyuki and Kwoen, Jinkwan and Arakawa, Yasuhiko and Hirakawa, Kazuhiko},
  journal = {Phys. Rev. Lett.},
  volume = {132},
  issue = {6},
  pages = {066901},
  numpages = {8},
  year = {2024},
  month = {Feb},
  publisher = {American Physical Society},
  doi = {10.1103/PhysRevLett.132.066901}
}

@article{Apretna2021,
title = {Few picosecond dynamics of intraband transitions in THz HgTe nanocrystals},
author = {Thibault Apretna and Sylvain Massabeau and Charlie Gréboval and Nicolas Goubet and Jérôme Tignon and Sukhdeep Dhillon and Francesca Carosella and Robson Ferreira and Emmanuel Lhuillier and Juliette Mangeney},
pages = {2753--2763},
volume = {10},
number = {10},
journal = {Nanophotonics},
doi = {10.1515/nanoph-2021-0249},
year = {2021},
lastchecked = {2024-11-12}
}

@article{Riccardi2020,
author = {Riccardi, Elisa and Massabeau, Sylvain and Valmorra, Federico and Messelot, Simon and Rosticher, Michael and Tignon, J{\'e}rôme and Watanabe, Kenji and Taniguchi, Takashi and Delbecq, Matthieu and Dhillon, Sukhdeep and Ferreira, Robson and Balibar, S{\'e}bastien and Kontos, Takis and Mangeney, Juliette},
title = {Ultrasensitive Photoresponse of Graphene Quantum Dots in the Coulomb Blockade Regime to THz Radiation},
journal = {Nano Lett.},
volume = {20},
number = {7},
pages = {5408-5414},
year = {2020},
doi = {10.1021/acs.nanolett.0c01800}
}

@article{Molitor_2010,
doi = {10.1209/0295-5075/89/67005},
year = {2010},
month = {apr},
publisher = {},
volume = {89},
number = {6},
pages = {67005},
author = {F. Molitor and H. Knowles and S. Dröscher and U. Gasser and T. Choi and P. Roulleau and J. Güttinger and A. Jacobsen and C. Stampfer and K. Ensslin and T. Ihn},
title = {Observation of excited states in a graphene double quantum dot},
journal = {Europhys. Lett.},
}

@article{Banszerus2020,
	author = {Banszerus, Luca and M{\"o}ller, Samuel and Icking, Eike and Watanabe, Kenji and Taniguchi, Takashi and Volk, Christian and Stampfer, Christoph},
	date = {2020/03/11},
	doi = {10.1021/acs.nanolett.9b05295},
	isbn = {1530-6984},
	journal = {Nano Lett.},
	journal1 = {Nano Lett.},
	journal2 = {Nano Lett.},
	month = {03},
	number = {3},
	pages = {2005--2011},
	publisher = {American Chemical Society},
	title = {Single-Electron Double Quantum Dots in Bilayer Graphene},
	type = {doi: 10.1021/acs.nanolett.9b05295},
	volume = {20},
	year = {2020},
    year1 = {2020}}

@article{Banszerus_tunable_coupling_BLG_QD,
    author = {Banszerus, L. and Rothstein, A. and Icking, E. and Möller, S. and Watanabe, K. and Taniguchi, T. and Stampfer, C. and Volk, C.},
    title = "{Tunable interdot coupling in few-electron bilayer graphene double quantum dots}",
    journal = {Appl. Phys. Lett.},
    volume = {118},
    number = {10},
    pages = {103101},
    year = {2021},
    month = {03},
    issn = {0003-6951},
    doi = {10.1063/5.0035300}
}

@article{Dani2018,
	author = {Dani, Olfa and Hussein, Robert and Bayer, Johannes C. and Kohler, Sigmund and Haug, Rolf J.},
	date = {2022/11/18},
	doi = {10.1038/s42005-022-01074-z},
	id = {Dani2022},
	isbn = {2399-3650},
	journal = {Commun. Phys.},
	number = {1},
	pages = {292},
	title = {Temperature-dependent broadening of coherent current peaks in InAs double quantum dots},
	volume = {5},
	year = {2022}
}

@article{Liu_2014,
  title = {Photon Emission from a Cavity-Coupled Double Quantum Dot},
  author = {Liu, Y.-Y. and Petersson, K. D. and Stehlik, J. and Taylor, J. M. and Petta, J. R.},
  journal = {Phys. Rev. Lett.},
  volume = {113},
  issue = {3},
  pages = {036801},
  numpages = {5},
  year = {2014},
  month = {Jul},
  publisher = {American Physical Society},
  doi = {10.1103/PhysRevLett.113.036801}
}

@article{Lundberg2024,
    author = {Lundberg, Theodor and Ibberson, David J. and Li, Jing and Hutin, Louis and Abadillo-Uriel, Jos{\'e}C. and Filippone, Michele and Bertrand, Benoit and Nunnenkamp, Andreas and Lee, Chang-Min and Stelmashenko, Nadia and Robinson, Jason W. A. and Vinet, Maud and Ibberson, Lisa and Niquet, Yann-Michel and Gonzalez-Zalba, M. Fernando},
	date = {2024/03/06},
	doi = {10.1038/s41534-024-00820-1},
	id = {Lundberg2024},
	isbn = {2056-6387},
	journal = {npj Quantum Inf.},
	number = {1},
	pages = {28},
	title = {Non-symmetric Pauli spin blockade in a silicon double quantum dot},
	volume = {10},
	year = {2024}
}

@article{Denisov_2022,
author = {Denisov, Artem O. and Oh, Seong W. and Fuchs, Gordian and Mills, Adam R. and Chen, Pengcheng and Anderson, Christopher R. and Gyure, Mark F. and Barnard, Arthur W. and Petta, Jason R.},
title = {Microwave-Frequency Scanning Gate Microscopy of a Si/SiGe Double Quantum Dot},
journal = {Nano Lett.},
volume = {22},
number = {12},
pages = {4807-4813},
year = {2022},
doi = {10.1021/acs.nanolett.2c01098}
}

@article{vanderWiel_2001,
doi = {10.1143/JJAP.40.2100},
year = {2001},
month = {mar},
publisher = {},
volume = {40},
number = {3S},
pages = {2100},
author = {Wilfred G. van der Wiel and Toshimasa Fujisawa and Seigo Tarucha and Leo P. Kouwenhoven},
title = {A Double Quantum Dot as an Artificial Two-Level System},
journal = {Jpn. J. Appl. Phys.},
}

@article{Neukelmance2025,
	author = {Neukelmance, B. and Hue, B. and Schaeverbeke, Q. and Jarjat, L. and Th{\'e}ry, A. and Craquelin, J. and Legrand, W. and Cubaynes, T. and Abulizi, G. and Becdelievre, J. and El Abbassi, M. and Larrouy, A. and Ourak, K. F. and Stefani, D. and Sulpizio, J. A. and Cottet, A. and Desjardins, M. M. and Kontos, T. and Delbecq, M. R.},
	journal = {Nat. Commun.},
	number = {1},
	pages = {5636},
	title = {Microsecond-lived quantum states in a carbon-based circuit driven by cavity photons},
	volume = {16},
	year = {2025},
    doi={10.1038/s41467-025-60952-6}}

@article{Zhang_2023,
doi = {10.1088/2633-4356/acfa57},
year = {2023},
month = {sep},
publisher = {IOP Publishing},
volume = {3},
number = {3},
pages = {036201},
author = {Jian Zhang and Liu Qian and Gabriela Borin Barin and Peipei Chen and Klaus Müllen and Pascal Ruffieux and Roman Fasel and Jin Zhang and Michel Calame and Mickael L Perrin},
title = {Double quantum dots in atomically-precise graphene nanoribbons},
journal = {Mater. Quantum Technol.},
}

@article{El_Fatimy2016,
   title={Epitaxial graphene quantum dots for high-performance terahertz bolometers},
   volume={11},
   ISSN={1748-3395},
   DOI={10.1038/nnano.2015.303},
   number={4},
   journal={Nat. Nanotechnol.},
   publisher={Springer Science and Business Media LLC},
   author={El Fatimy, Abdel and Myers-Ward, Rachael L. and Boyd, Anthony K. and Daniels, Kevin M. and Gaskill, D. Kurt and Barbara, Paola},
   year={2016},
   month=jan, pages={335–338}}

@article{Messelot2022,
  title = {Large terahertz electric dipole of a single graphene quantum dot},
  author = {Messelot, Simon and Riccardi, Elisa and Massabeau, Sylvain and Valmorra, Federico and Rosticher, Michael and Watanabe, Kenji and Taniguchi, Takashi and Tignon, J\'er\^ome and Boulier, Thomas and Dauvois, Vincent and Delbecq, Matthieu and Dhillon, Sukhdeep and Balibar, S\'ebastien and Kontos, Takis and Mangeney, Juliette},
  journal = {Phys. Rev. Res.},
  volume = {4},
  issue = {1},
  pages = {L012018},
  numpages = {6},
  year = {2022},
  month = {Feb},
  publisher = {American Physical Society},
  doi = {10.1103/PhysRevResearch.4.L012018}
}

@article{Wang2013,
author = {L. Wang  and I. Meric  and P. Y. Huang  and Q. Gao  and Y. Gao  and H. Tran  and T. Taniguchi  and K. Watanabe  and L. M. Campos  and D. A. Muller  and J. Guo  and P. Kim  and J. Hone  and K. L. Shepard  and C. R. Dean },
title = {One-Dimensional Electrical Contact to a Two-Dimensional Material},
journal = {Science},
volume = {342},
number = {6158},
pages = {614-617},
year = {2013},
doi = {10.1126/science.1244358}
}

@article{Banszerus2015,
author = {Luca Banszerus  and Michael Schmitz  and Stephan Engels  and Jan Dauber  and Martin Oellers  and Federica Haupt  and Kenji Watanabe  and Takashi Taniguchi  and Bernd Beschoten  and Christoph Stampfer },
title = {Ultrahigh-mobility graphene devices from chemical vapor deposition on reusable copper},
journal = {Sci. Adv.},
volume = {1},
number = {6},
pages = {e1500222},
year = {2015},
doi = {10.1126/sciadv.1500222}
}

@article{Guttinger2012,
doi = {10.1088/0034-4885/75/12/126502},
year = {2012},
month = {nov},
publisher = {IOP Publishing},
volume = {75},
number = {12},
pages = {126502},
author = {J Güttinger and F Molitor and C Stampfer and S Schnez and A Jacobsen and S Dröscher and T Ihn and K Ensslin},
title = {Transport through graphene quantum dots},
journal = {Rep. Prog. Phys.},
}

@article{Liu2010,
author = {Liu, Xing Lan and Hug, Dorothee and Vandersypen, Lieven M. K.},
title = {Gate-defined graphene double quantum dot and excited state spectroscopy},
journal = {Nano Lett.},
volume = {10},
number = {5},
pages = {1623-1627},
year = {2010},
doi = {10.1021/nl9040912}
}

@article{vanderWiel2002,
  title = {Electron transport through double quantum dots},
  author = {van der Wiel, W. G. and De Franceschi, S. and Elzerman, J. M. and Fujisawa, T. and Tarucha, S. and Kouwenhoven, L. P.},
  journal = {Rev. Mod. Phys.},
  volume = {75},
  issue = {1},
  pages = {1--22},
  numpages = {0},
  year = {2002},
  month = {Dec},
  publisher = {American Physical Society},
  doi = {10.1103/RevModPhys.75.1}
}

@article{Nill2024,
  title = {Avalanche Terahertz Photon Detection in a Rydberg Tweezer Array},
  author = {Nill, Chris and Cabot, Albert and Trautmann, Arno and Gro\ss{}, Christian and Lesanovsky, Igor},
  journal = {Phys. Rev. Lett.},
  volume = {133},
  issue = {7},
  pages = {073603},
  numpages = {7},
  year = {2024},
  month = {Aug},
  publisher = {American Physical Society},
  doi = {10.1103/PhysRevLett.133.073603}
}

@article{Wade2017,
	author = {Wade, C. G. and {\v S}ibali{\'c}, N. and de Melo, N. R. and Kondo, J. M. and Adams, C. S. and Weatherill, K. J.},
	date = {2017/01/01},
	doi = {10.1038/nphoton.2016.214},
	id = {Wade2017},
	isbn = {1749-4893},
	journal = {Nat. Photonics},
	number = {1},
	pages = {40--43},
	title = {Real-time near-field terahertz imaging with atomic optical fluorescence},
	volume = {11},
	year = {2017}}

@article{Nazarov_1993,
title = {Quantum interference, tunnel junctions and resonant tunneling interferometer},
journal = {Physica B: Condensed Matter},
volume = {189},
number = {1},
pages = {57-69},
year = {1993},
issn = {0921-4526},
doi = {10.1016/0921-4526(93)90146-W},
author = {Yuli V. Nazarov},
}

@article{Stoof_Nazarov_1996,
   title={Time-dependent resonant tunneling via two discrete states},
   volume={53},
   ISSN={1095-3795},
   DOI={10.1103/physrevb.53.1050},
   number={3},
   journal={Phys. Rev. B},
   publisher={American Physical Society (APS)},
   author={Stoof, T. H. and Nazarov, Yu. V.},
   year={1996},
   month=jan, pages={1050–1053} }

@article{Sztenkiel_2007,
  title   = {Electron Transport through Double Quantum Dot System with Inter-Dot Coulomb Interaction},
  author  = {Sztenkiel, D. and {\'S}wirkowicz, R.},
  journal = {Acta Phys Pol. A},
  volume  = {111},
  number  = {3},
  pages   = {361--372},
  year    = {2007},
  doi     = {10.12693/APhysPolA.111.361}
}

@article{Bulka2004,
  title = {Electronic correlations in coherent transport through a two quantum dot system},
  author = {Bu\l{}ka, Bogdan R. and Kostyrko, Tomasz},
  journal = {Phys. Rev. B},
  volume = {70},
  issue = {20},
  pages = {205333},
  numpages = {11},
  year = {2004},
  month = {Nov},
  publisher = {American Physical Society},
  doi = {10.1103/PhysRevB.70.205333}
}

@article{Sobrino2024,
  title = {Fully analytical equation of motion approach for the double quantum dot in the Coulomb blockade regime},
  author = {Sobrino, Nahual and Jacob, David and Kurth, Stefan},
  journal = {Phys. Rev. B},
  volume = {110},
  issue = {11},
  pages = {115121},
  numpages = {13},
  year = {2024},
  month = {Sep},
  publisher = {American Physical Society},
  doi = {10.1103/PhysRevB.110.115121}
}

@article{Meir1992,
  title = {Landauer formula for the current through an interacting electron region},
  author = {Meir, Yigal and Wingreen, Ned S.},
  journal = {Phys. Rev. Lett.},
  volume = {68},
  issue = {16},
  pages = {2512--2515},
  numpages = {0},
  year = {1992},
  month = {Apr},
  publisher = {American Physical Society},
  doi = {10.1103/PhysRevLett.68.2512}
}

@article{Lamba2000,
  title = {Transport through a coupled quantum dot system: Role of interdot interactions},
  author = {Lamba, Sushil and Joshi, S. K.},
  journal = {Phys. Rev. B},
  volume = {62},
  issue = {3},
  pages = {1580--1583},
  numpages = {0},
  year = {2000},
  month = {Jul},
  publisher = {American Physical Society},
  doi = {10.1103/PhysRevB.62.1580}
}

@article{You1999,
  title = {Spectral properties of a double-quantum-dot structure: A causal Green's function approach},
  author = {You, J. Q. and Zheng, Hou-Zhi},
  journal = {Phys. Rev. B},
  volume = {60},
  issue = {12},
  pages = {8727--8733},
  numpages = {0},
  year = {1999},
  month = {Sep},
  publisher = {American Physical Society},
  doi = {10.1103/PhysRevB.60.8727}
}

@article{Crouch1997,
    author = {Crouch, C. H. and Livermore, C. and Westervelt, R. M. and Campman, K. L. and Gossard, A. C.},
    title = {Evolution of the Coulomb gap in tunnel-coupled quantum dots},
    journal = {Appl. Phys. Lett.},
    volume = {71},
    number = {6},
    pages = {817-819},
    year = {1997},
    month = {08},

    issn = {0003-6951},
    doi = {10.1063/1.119656}
}

@article{Gustavsson2008,
  title = {Detecting single-electron tunneling involving virtual processes in real time},
  author = {Gustavsson, S. and Studer, M. and Leturcq, R. and Ihn, T. and Ensslin, K. and Driscoll, D. C. and Gossard, A. C.},
  journal = {Phys. Rev. B},
  volume = {78},
  issue = {15},
  pages = {155309},
  numpages = {9},
  year = {2008},
  month = {Oct},
  publisher = {American Physical Society},
  doi = {10.1103/PhysRevB.78.155309}
}

@article{
Livermore1996,
author = {C. Livermore  and C. H. Crouch  and R. M. Westervelt  and K. L. Campman  and A. C. Gossard },
title = {The Coulomb Blockade in Coupled Quantum Dots},
journal = {Science},
volume = {274},
number = {5291},
pages = {1332-1335},
year = {1996},
doi = {10.1126/science.274.5291.1332}
}

@article{Kasprzak2010,
	author = {Kasprzak, J. and Reitzenstein, S. and Muljarov, E. A. and Kistner, C. and Schneider, C. and Strauss, M. and H{\"o}fling, S. and Forchel, A. and Langbein, W.},
	date = {2010/04/01},
	doi = {10.1038/nmat2717},
	id = {Kasprzak2010},
	isbn = {1476-4660},
	journal = {Nat. Mater.},
	number = {4},
	pages = {304--308},
	title = {Up on the Jaynes--Cummings ladder of a quantum-dot/microcavity system},
	volume = {9},
	year = {2010}}

@article{Hecker2023,
	author = {Hecker, K. and Banszerus, L. and Sch{\"a}pers, A. and M{\"o}ller, S. and Peters, A. and Icking, E. and Watanabe, K. and Taniguchi, T. and Volk, C. and Stampfer, C.},
	date = {2023/11/30},
	doi = {10.1038/s41467-023-43541-3},
	id = {Hecker2023},
	isbn = {2041-1723},
	journal = {Nat. Commun.},
	number = {1},
	pages = {7911},
	title = {Coherent charge oscillations in a bilayer graphene double quantum dot},
	volume = {14},
	year = {2023}}

@article{
Petta2005,
author = {J. R. Petta  and A. C. Johnson  and J. M. Taylor  and E. A. Laird  and A. Yacoby  and M. D. Lukin  and C. M. Marcus  and M. P. Hanson  and A. C. Gossard },
title = {Coherent Manipulation of Coupled Electron Spins in Semiconductor Quantum Dots},
journal = {Science},
volume = {309},
number = {5744},
pages = {2180-2184},
year = {2005},
doi = {10.1126/science.1116955}
}

@article{Kouwenhoven1999,
title = {Microwave spectroscopy of a double quantum dot in the low- and high-power regime},
journal = {Physica B: Condensed Matter},
volume = {272},
number = {1},
pages = {31-35},
year = {1999},
issn = {0921-4526},
doi = {10.1016/S0921-4526(99)00246-X},
author = {W.G {van der Wiel} and T Fujisawa and T.H Oosterkamp and L.P Kouwenhoven},
}

@article{
Fujisawa1998,
author = {Toshimasa Fujisawa  and Tjerk H. Oosterkamp  and Wilfred G. van der Wiel  and Benno W. Broer  and Ramón Aguado  and Seigo Tarucha  and Leo P. Kouwenhoven },
title = {Spontaneous Emission Spectrum in Double Quantum Dot Devices},
journal = {Science},
volume = {282},
number = {5390},
pages = {932-935},
year = {1998},
doi = {10.1126/science.282.5390.932}
}

@article{vanderVaart1995,
  title = {Resonant Tunneling Through Two Discrete Energy States},
  author = {van der Vaart, N. C. and Godijn, S. F. and Nazarov, Y. V. and Harmans, C. J. P. M. and Mooij, J. E. and Molenkamp, L. W. and Foxon, C. T.},
  journal = {Phys. Rev. Lett.},
  volume = {74},
  issue = {23},
  pages = {4702--4705},
  numpages = {0},
  year = {1995},
  month = {Jun},
  publisher = {American Physical Society},
  doi = {10.1103/PhysRevLett.74.4702}
}

@article{Navarro2016,
	author = {Navarro-C{\'\i}a, Miguel and Wu, Jiang and Liu, Huiyun and Mitrofanov, Oleg},
	date = {2016/12/12},
	doi = {10.1038/srep38926},
	id = {Navarro-C{\'\i}a2016},
	isbn = {2045-2322},
	journal = {Sci. Rep.},
	number = {1},
	pages = {38926},
	title = {Generation of radially-polarized terahertz pulses for coupling into coaxial waveguides},
	volume = {6},
	year = {2016}}

@article{Tarucha1999,
title = {Elastic and inelastic single electron tunneling in coupled two dot system},
journal = {Microelectron. Eng.},
volume = {47},
number = {1},
pages = {101-105},
year = {1999},
issn = {0167-9317},
doi = {10.1016/S0167-9317(99)00162-8},
author = {Seigo Tarucha and T. Fujisawa and K. Ono and D.G. Austing and T.H. Oosterkamp and W.G. {van der Wiel} and L.P. Kouwenhoven}
}

@article{Kellert2017,
  title = {Few-Electron Ultrastrong Light-Matter Coupling at 300 GHz with Nanogap Hybrid LC Microcavities},
  author = {Keller, J. and Scalari, G. and Cibella, S. and Maissen, C. and Appugliese, F. and Giovine, E. and Leoni, R. and Beck, M. and Faist, J.},
  journal = {Nano Lett.},
  volume = {17},
  issue = {12},
  pages = {7410–7415},
  year = {2017},
  publisher = {American Chemical Society},
  doi={10.1021/acs.nanolett.7b03228}
}

@article{Pontula2025,
title = {Strong broadband intensity noise squeezing from infrared to terahertz frequencies in lasers with nonlinear dissipation},
author = {Sahil Pontula and Jamison Sloan and Nicholas Rivera and Marin Soljačić},
pages = {3243--3254},
volume = {14},
number = {20},
journal = {Nanophotonics},
doi = {10.1515/nanoph-2025-0259},
year = {2025},
lastchecked = {2025-12-23}
}

@article{Zhang2025,
	author = {Zhang, Jingda and Quek, Su Ying},
	journal = {ACS Nano},
	month = {10},
	number = {41},
	pages = {36204--36214},
	title = {Quantum Defects in 2D Transition Metal Dichalcogenides for Terahertz Technologies},
	volume = {19},
	year = {2025},
    doi={10.1021/acsnano.5c06007}}

@article{Messelot2023,
author = {Simon Messelot and Solen Coeymans and J\'{e}r\^{o}me Tignon and Sukhdeep Dhillon and Juliette Mangeney},
journal = {Photon. Res.},
number = {7},
pages = {1203--1216},
publisher = {Optica Publishing Group},
title = {High Q and sub-wavelength THz electric field confinement in ultrastrongly coupled THz resonators},
volume = {11},
month = {Jul},
year = {2023},
doi = {10.1364/PRJ.482195},
}

@article{Baydin2025,
author = {Andrey Baydin and Hanyu Zhu and Motoaki Bamba and Kaden R. A. Hazzard and Junichiro Kono},
journal = {Opt. Mater. Express},
number = {8},
pages = {1833--1846},
publisher = {Optica Publishing Group},
title = {Perspective on the quantum vacuum in matter},
volume = {15},
month = {Aug},
year = {2025},
doi = {10.1364/OME.569339},
}

@misc{Johl2025,
      title={Gate-tunable single terahertz meta-atom ultrastrong light-matter coupling}, 
      author={Elsa Jöchl and Anna-Lydia Vieli and Lucy Hale and Felix Helmrich and Deniz Turan and Mona Jarrahi and Mattias Beck and Jérôme Faist and Giacomo Scalari},
      year={2025},
      eprint={2511.03664},
      archivePrefix={arXiv},
      primaryClass={cond-mat.mes-hall},
      url={https://arxiv.org/abs/2511.03664}, 
}

@article{Mirco2025,
author = {Kutas, Mirco and Riexinger, Felix and Klier, Jens and Molter, Daniel and von Freymann, Georg},
title = {Terahertz Quantum Imaging},
journal = {Adv. Quantum Technol.},
volume = {8},
number = {11},
pages = {e2500271},
doi = {10.1002/qute.202500271},
year = {2025}
}

@article{Liu2025,
	author = {Liu, Mingqing and Kazemi, Hossein and Safari, Majid and Tavakkolnia, Iman and Haas, Harald},
	journal = {npj Wirel. Technol.},
	number = {1},
	pages = {3},
	title = {A comprehensive comparison between Terahertz and optical wireless communications},
	volume = {1},
	year = {2025},
    doi={10.1038/s44459-025-00002-1}}

@article{Novoselov2016,
author = {K. S. Novoselov  and A. Mishchenko  and A. Carvalho  and A. H. Castro Neto },
title = {2D materials and van der Waals heterostructures},
journal = {Science},
volume = {353},
number = {6298},
pages = {aac9439},
year = {2016},
doi = {10.1126/science.aac9439}
}

@article{cole2001,
  title={Coherent manipulation of semiconductor quantum bits with terahertz radiation},
  author={B. E. Cole and J. B. Williams and B. T. King and M. S. Sherwin and C. R. Stanley},
  journal={Nature},
  volume={410},
  number={6824},
  pages={60--63},
  year={2001},
  publisher={Nature Publishing Group UK London},
  doi={10.1038/35065032}}

@article{wang2012,
  title={Controllable tunnel coupling and molecular states in a graphene double quantum dot},
  author={Wang, Lin-Jun and Li, Hai-Ou and Tu, Tao and Cao, Gang and Zhou, Cheng and Hao, Xiao-Jie and Su, Zhan and Xiao, Ming and Guo, Guang-Can and Chang, Albert M and Guo, Guo-Ping},
  journal={Appl. Phys. Lett.},
  volume={100},
  number={2},
  pages={022106},
  year={2012},
  publisher={AIP Publishing},
  doi = {10.1063/1.3676083}
}

@article{Engels2013,
    author = {Engels, S. and Epping, A. and Volk, C. and Korte, S. and Voigtländer, B. and Watanabe, K. and Taniguchi, T. and Trellenkamp, S. and Stampfer, C.},
    title = {Etched graphene quantum dots on hexagonal boron nitride},
    journal = {Appl. Phys. Lett.},
    volume = {103},
    number = {7},
    pages = {073113},
    year = {2013},
    month = {08},
    issn = {0003-6951},
    doi = {10.1063/1.4818627}
}
\bibliographystyle{sciencemag}

%%%%%%%%%%%%%%%% ACKNOWLEDGEMENTS %%%%%%%%%%%%%%%

\paragraph*{Acknowledgments:}
We are grateful to Z. Leghtas, B. Kuhlmey, I. Abdourahamane, S. Massabeau and R. Ferreira for fruitful discussions.

\paragraph*{Funding:}
The work in LPENS was supported by French State Funds managed by PEPR Electronique, the ANR using ThinQ (No. ANR-23-GRF1-0002), QuickTera (No. ANR-22-CE09-0018), MicroTera (No. ANR-24-CE30-0621), and the DIM Quantip using TeraQD (No. 276133). K.W. and T.T. acknowledge support from the JSPS KAKENHI (Grant Numbers 21H05233 and 23H02052) and World Premier International Research Center Initiative (WPI), MEXT, Japan.

\paragraph*{Author contributions:} 

Conceptualization: J.M. 
Formal analysis: A.W., A.C., T.K., J.M. 
Funding acquisition: J.M. 
Methodology: A.W., S.C., T.K., J.M. 
Investigation: A.W., S.C., S.M., J.M. 
Resources: A.S., M.R., K.W., T.T., J.T.
Software: A.W., S.C. 
Visualization: A.W. 
Writing - original draft: A.W., M.D., T.K., J.M. 
Writing - review \& editing: A.W., B.H., S.D., R.R., A.C., M.D., T.K., J.M. 
Supervision: J.M.

\paragraph*{Competing interests:}
All authors declare they have no competing interests.

\paragraph*{Data and materials availability:}
All data and code needed to evaluate and reproduce the results in the paper are present in the paper and/or the Supplementary Materials. This study did not generate new materials.

% on request not allowed

%Cite the relevant DOI \cite{dataset}, URL \cite{example_url} or reference \cite{example2} in this statement. These \textit{do not} count towards the reference limit if they are only cited in the acknowledgements. Be specific and state a unique identifier -- such as an accession number, software version number or observation ID -- so readers can easily retrieve the exact material used.

%%%%%%%%%%%%%%%% SUPPLEMENT LIST %%%%%%%%%%%%%%%

% List the contents of your Supplementary Materials, including the numbers of any
% supplementary figures, tables, external data files etc. and any references that are
% cited only in the supplement. In this example, refs. 7-8 are cited only in the supplement.
% Fill out your numbers accordingly and delete any lines that aren't applicable.

\end{document}